\documentclass[runningheads]{llncs}
\usepackage{multirow}
\usepackage[T1]{fontenc}
\usepackage[utf8]{inputenc}
\usepackage{graphicx}
\usepackage{wrapfig}
\usepackage{subcaption}
\usepackage{xcolor}
\usepackage[most]{tcolorbox}
\usepackage{tikz}
\usetikzlibrary{arrows.meta}
\usetikzlibrary{positioning, patterns, calc, shapes, arrows, shapes.misc}
\usetikzlibrary{decorations.pathreplacing, shapes.multipart}
\tikzstyle{neuron}=[draw,circle,thick,inner sep=0, minimum size = 20]
\tikzstyle{cfgnode}=[draw,circle,thick,inner sep=5,
minimum width = 30, blue, fill=blue!5, text=black, align=left]
\tikzstyle{inactivation} = [cross out, draw=red, line width=0.75mm,
minimum size = 10pt, inner sep=0pt, outer sep=0pt]
\tikzstyle{activation} = [circle, draw=olive, line width=0.75mm,
minimum size = 10pt]
\tikzstyle{wv} = [line width=0.75mm, midway, above]
\tikzstyle{modulesingle} = [rectangle, draw, thick, align=center,
inner sep=0.2cm,
orange, text=black, fill = orange!20, rounded corners]
\usetikzlibrary{calc}
\usepackage{relsize}

\tikzset{fontscale/.style = {font=\relsize{#1}}
    }

\usepackage{booktabs}
\usepackage{amsmath}
\usepackage{amsfonts}
\usepackage{amssymb}
\usepackage{amssymb}
\usepackage{subcaption}
\usepackage{mathtools}
\usepackage{comment}
\usepackage{url}
\usepackage{xspace}
\usepackage{listings}
\usepackage{bbm}
\newcommand{\hypothesis}{\mathcal{H}}
\newcommand{\problemSpace}{\mathcal{P}}
\newcommand{\resultSpace}{\mathcal{R}}

\newcommand{\system}{s}
\newcommand{\systemProperty}{\Psi}
\newcommand{\solutionProperty}{\Phi}
\newcommand{\networkProperty}{\Xi}

\newcommand{\R}{\mathbb{R}}

\newcommand{\Ereal}[0]{[-\infty,\infty]}

\usepackage{caption}
\usepackage{todonotes}

\begin{document}
\title{Proof-Carrying Neuro-Symbolic Code}
\titlerunning{Proof-Carrying Neuro-Symbolic Code}

\author{ Ekaterina Komendantskaya}

\institute{Southampton University, UK \and Heriot-Watt University, UK }

\authorrunning{E. Komendantskaya}

\maketitle              
\begin{abstract}
This invited paper introduces the concept of  ``proof-carrying neuro-symbolic code'' and explains its meaning and value, from both the ``neural'' and the ``symbolic'' perspectives. The talk outlines the first successes and challenges that this new area of research faces.    

\end{abstract}

\section{Neuro-Symbolic Proofs and Programs}

\emph{Proof Carrying Code} is a long tradition within programming language research, broadly referring to methods that 
interleave verification with executable code, thus avoiding the inevitable discrepancies 
that arise when the code and the proofs are handled in different languages. Although the term was coined by Necula~\cite{Necula97} almost three decades ago, with time,
it grew to encompass any languages that are powerful enough to handle both the coding and the proving. Examples are dependently-typed (Agda, Idris, Coq/Rocq) and refinement-typed (F*, Liquid Haskell) languages. In the last decades,  both families of languages have seen substantial successes in the proof-carrying code direction~\cite{10.1145/3607844,leroyFormalVerificationRealistic2009}.

 An equally impressive, though very different in nature, revolution has happened elsewhere in computer science: machine learning methods grew in quality, diversity and quickly proliferated to different applications.  Machine learning attracted attention of programming language researchers, and there have been recent advances in functional algorithms for automated differentiation, probabilistic programming, as well as \emph{neuro-symbolic programming} and \emph{neuro-symbolic theorem proving}. 
The latter two aim to develop, respectively:
\begin{itemize}
\item methods of merging machine learning code and standard (symbolic) code~\cite{PGL-049}; 

and  

\item methods automating theorem proving with the help of machine learning deployed at the stage of proof search or lemma conjecturing~\cite{DBLP:journals/corr/abs-1212-3618,DBLP:conf/lpar/HerasKJM13,DBLP:journals/corr/abs-2412-16075}.
\end{itemize}

    The natural question to ask is: \emph{Does the concept of proof-carrying code bear any value, or indeed meaning, in the age of the
     neuro-symbolic paradigm shift}? This invited talk will answer both questions in the positive. 

\section{What is Proof-Carrying Neuro-Symbolic Code?}\label{sec:nesy}
\begin{figure}[t]
\begin{center}
\includegraphics[width=0.7\textwidth,height=1.7cm]{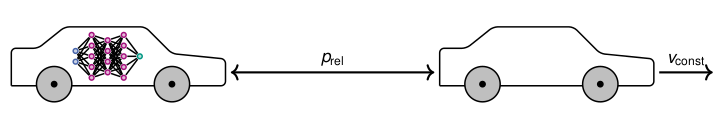}
\end{center}
\footnotesize{$$ p_{rel} > \frac{v_{rel}^2}{2B} \rightarrow  \bigl[{\color{red}a_{rel}} := \textcolor{red}{f}(\textcolor{blue}{v_{rel},p_{rel}}); t := 0 ; {\color{blue}\{ p'_{rel} = v_{rel}, v'_{rel} = -a_{rel}, t' = 1 \}}\bigr] p_{rel} >0$$}
  \caption{\footnotesize{\emph{Example of a cyber-physical system with a neural controller (given by the left-most car) written in the syntax of the Differentiable Dynamic Logic~\cite{FultonMQVP15,Platzer18,TeuberMP24}. 
  }}}
 \label{fig:cars}
\end{figure}
Consider a purely symbolic program $\system(\cdot)$, whose completion requires computing a complex, unknown function~$\hypothesis : \problemSpace \rightarrow \resultSpace$ that maps objects in the \emph{problem input space} $\problemSpace$ to those in the \emph{problem output space}~$\resultSpace$. Given an embedding function $e : \problemSpace \rightarrow \R^m$ and an unembedding function $u : \R^n \rightarrow \resultSpace$, we can approximate~$\hypothesis$ by training a neural network~${f : \R^m \rightarrow \R^n}$ such that $u \circ f \circ e \approx \hypothesis$. We refer to $u \circ f \circ e$ as the \emph{solution}, and refer to $\R^m$ and $\R^n$ as the \emph{embedding input space} and \emph{embedding output space} respectively. Unlike objects in the problem space, the vectors in the embedding space are often not directly interpretable. The complete \emph{neuro-symbolic program} is then modelled as $\system(u \circ f \circ e)$. 
Examples of $u$ and $e$ would be the normalization of inputs, resizing operations for images, or data augmentation operations that are commonplace in machine learning pipelines. Most common examples of neuro-symbolic programs come from the area of \emph{cyber-physical system} modelling, as such systems often implement machine learning algorithms in order to process sensor data. We will illustrate this class of applications using a simple example (Example~\ref{ex:cars} and Fig.~\ref{fig:cars}).

What might one want to prove about such a neuro-symbolic program? The end goal may be to prove that~$\system(u \circ f \circ e)$ satisfies a property~$\systemProperty$, which we will refer to as the \emph{program property}. The natural way to proceed is to establish a \emph{solution property}~$\solutionProperty$ and a \emph{network property}~$\networkProperty$ such that the proof of~$\systemProperty$ is decomposable into the following three lemmas:
\begin{align}
    \label{eq:networkProperty}
    \networkProperty(f)
    \qquad \qquad \qquad \qquad \qquad \qquad \qquad \qquad \qquad \qquad \\
    \label{eq:solutionProperty}
    \forall g : \networkProperty(g) \Rightarrow \solutionProperty(u \circ g \circ e)
    \qquad \qquad \qquad \qquad \quad                                     \\ \qquad \qquad \qquad \qquad
    \label{eq:systemProperty}
    \forall h : \solutionProperty(h) \Rightarrow
    \systemProperty(\system(h))
\end{align}
i.e. Lemma~\ref{eq:networkProperty} proves that the network~$f$ obeys the network property~$\networkProperty$, then Lemma~\ref{eq:solutionProperty} proves that this implies $u \circ f \circ e$, the neural network lifted to the problem space, obeys the solution property~$\solutionProperty$, and finally Lemma~\ref{eq:systemProperty} proves that this implies $\system(u \circ f \circ e)$, the neuro-symbolic program, obeys the program property~$\systemProperty$.

 \emph{In conclusion, writing proof-carrying neuro-symbolic code  amounts to writing a program $\system(u \circ f \circ e)$ and completing a proof as in equations (1) -- (3)}.  

\begin{example}[A Cyber-Physical System with a Neural Network Controller]\label{ex:cars}
Fig.~\ref{fig:cars} shows to cars moving. The second car is ``autonomous", i.e. it has a neural network \textcolor{red}{$f$} for controlling the \textcolor{red}{(de)acceleration $a_{rel}$} (we use red colour to indicate that it is potentially unsafe). 
  The high-level safety specification boils down to asserting that, 
\emph{assuming the front car has constant speed, the back car can be trusted to 
maintain a safe distance}.
The \textcolor{blue}{relative distance $p_{rel}$} and \textcolor{blue}{velocity $v_{rel}$} are modelled by differential equations. We highlight the dynamical (continuous) parts of the system with blue colour.
 The \textcolor{blue}{relative distance $p_{rel}$} between the two cars is a function of velocity of the autonomous car (\textcolor{blue}{$p'_{rel} = v_{rel}$}), and the latter  
is a function of the car's \textcolor{blue}{relative acceleration: $v'_{rel} = -a_{rel}$}. 
This continuous dynamics has to interact with the discrete actions of the autonomous car; e.g., ``set the acceleration to value $\textcolor{red}{ f}(\textcolor{blue}{v_{rel},p_{rel}})$", written as $\textcolor{red}{a_{rel}} := \textcolor{red}{f}(\textcolor{blue}{v_{rel},p_{rel}})$, or ``start following the car in front of you at time $0$" ($t := 0$). 
Traditionally, the simplest way to control this car was to set $a_{rel} := -B$, for some breaking constant $B$. The neural network $\textcolor{red}{f}$ replaces this classical discrete controller. In the classical example,   
the safety specification is given by pre- and post-conditions: assuming the autonomous car satisfies $p_{rel} > \frac{v_{rel}^2}{2B}$, there will never be a collision: $p_{rel} > 0$. 
Suppose a neural controller \textcolor{red}{$f$} is trained on previous data, the  input-output pairs where the input is given by sensor readings  \textcolor{blue}{$v_{rel}$, $p_{rel}$} and the output by $-B$.\footnote{
This training scenario is somewhat convoluted, as we are assuming that the neural network is trained to simulate the classical controller which in this toy example was given by $a:= -B$, but the reader should imagine that in a real-life examples the training data comes from a less trivial distribution.}
For training prposes, the data is normalised as floating point numbers between some fixed interval, e.g. $[-1,1]$, thus \textcolor{red}{$f:[-1,1]^2 \rightarrow [-1,1]$}. In order for the whole cyber-physical system to be proven safe, the neural network $\textcolor{red}{f}$ must be verified to satisfy the following property (subject to normalisation of input and outputs of $\textcolor{red}{f}):   p_{rel} > \frac{v_{rel}^2}{2B}\rightarrow \textcolor{red}{f}(\textcolor{blue}{p_{rel},v_{rel}}) \leq -B$. This will then provably imply that the desired postcondition $p_{rel} > 0$ holds. 
\end{example}

The purpose of \emph{proof-carrying neuro-symbolic code} is to propose languages or programming environments where implementation of such a system, as well as all necessary proofs of its safety, can be carried out in a sound and consistent way. This has proven to be a challenge, see~\cite{PLNNV25} for a detailed discussion of the problems raised in the literature.


\section{Why Proof-Carrying Neuro-Symbolic Code?}\label{sec:ML}
Arguably, there are good reasons for both machine learning and programming language communities to invest efforts into the development of  proof-carrying neuro-symbolic code.

\textbf{Machine Learning Perspective.}
Traditionally, statistical machine learning has distinguished its methods from ``algorithm-driven'' programming: the consensus has been that machine learning is deployed when there is example input-output data but no general algorithm for computing outputs from inputs. Thus, neural networks are commonly seen as programs that emerge from data via training, without direct human guidance on how to perform the computation. This unfortunate dichotomy has led to a divide between programming language and machine learning research that is still awaiting resolution. 

The first hint that this dichotomy is not as fundamental as was thought came from the machine learning community itself. The famous paper by Szegedy et al.~\cite{szegedy2014intriguing} pointed out the ``intriguing'' problem that even the most accurate neural networks fail to satisfy the property of \emph{robustness}, i.e. small perturbations of their inputs should result in small changes to their output. Szegedy's key example concerned imperceptible perturbations of pixels in an image that can sway the neural network's classification decisions. This lack of robustness can have safety and security implications: for example, an autonomous car's vision unit may fail to recognise pedestrians on the road. For that reason, the problem attracted significant attention~\cite{Carlini19} but remains unresolved to this day.
Partial solutions often deploy methods of \emph{adversarial training} --- i.e., training based on computing \emph{adversarial attacks} --- which augment the training set with the worst-case perturbations of the input data points with respect to the output loss of the neural network~\cite{KM18}.

The robustness of neural networks actually yields a formal specification~\cite{CKDKKAE22}. Given a neural network $f: \R^m \to \R^n$, $f$ is \emph{robust around $\hat{x} \in \R^m$}, if 
 \begin{align}
 \label{robustness}  
 \forall x, \| \hat{x} - x\| \leq \epsilon  \implies  \| f(\hat{x}) - f(x)\| \leq \delta ,   
 \end{align}
 
\noindent where $\epsilon, \delta \in \R$ are small constants and $\|.\|$ computes a vector distance. From the programming language perspective, robustness can be seen as a refinement type that refines input and output types of $f$, cf.~\cite{KKKAA20}.
%
At the same time, robustness is an example of a \emph{desirable property} that neural networks cannot
learn from data alone: note the quantification over vectors $x$ that do not belong to the data set. 
This challenges the classical dichotomy between algorithm-driven and data-driven programming, demonstrating the inevitability of property specification in both cases -- and the need for solutions that are \emph{proof-carrying}.

\textbf{Proofs and Programs Perspective.} 
It is clear that neuro-symbolic systems generally, and cyber-physical systems in particular, are the next holy grail for  formal verification. This gives one reason for proof and program communities to engage in designing the principles 
 of proof-carrying neuro-symbolic code.
The less obvious reason, that will be demonstrated in the next section, is that, despite seemingly exotic nature of the neuro-symbolic set up, some of the best methods developed in the programming language community in the past few decades find unexpectedly direct applications in this new domain.

\section{Challenges in Developing Proof-Carrying Neuro-Symbolic Code}



\subsection{Challenge 1.  Multi-backend Interfaces}
The languages rich enough to host proof-carrying code generally (e.g. Agda, Rocq or Idris), and prove higher-level properties such as (\ref{eq:systemProperty}) in particular, do not tend to be good at proving low-level \emph{reachability properties} of functions over real vector spaces (cf.   $\networkProperty(f)$  in Eq. (1) or Eq. (4) in Section~\ref{sec:ML}). There exists a range of specialised solvers that can prove properties such as $\networkProperty(f)$, e.g.
Marabou~\cite{katz2019marabou,Wu24}, $\alpha\beta$-CROWN~\cite{wang2021beta} and PyRAT~\cite{pyratanalyzer2024PyRAT}.
These solvers have a specialised syntax to handle proof search for properties such as $\networkProperty(f)$. But this syntax is not general enough to serve as a proof-carrying code for general properties such $\systemProperty(\system(h))$. 

In many ways, the argument replicates the one that has already been made by programming languages such as F* and Liquid Haskell that successfully argued for delegation of low-level proofs to the SMT solvers. 
The language \textbf{Vehicle}~\cite{DaggittKKA0SCCL23,DBLP:journals/corr/abs-2401-06379,Vehicle} was the first to adapt this argument to neural network solvers. The adaptation  introduced a few interesting ideas concerning type checking and compilation fit for the neuro-symbolic set up~\cite{daggitt2024efficientcompilationexpressiveproblem,daggitt2023compiling}.

\textbf{Vehicle} provides an interface and a DSL (\emph{Domain-Specific Language}) in which properties such as (2) can be written, type checked, and compiled to both low-level solvers that can prove the property (1) and high-level languages that can express the property (3).  At the time of writing, however, \textbf{Vehicle} has only interface to Marabou for part (1) and to Agda for part (3) of the pipeline shown in Section~\ref{sec:nesy}. For practical applications, the range of solvers and theorem provers to which \textbf{Vehicle} compiles to needs to grow.  Crucially, it must involve higher-order  provers that are better geared to work with features that are valuable in cyber-physical system verification, for example:
\begin{itemize}
\item proving properties involving systems of ODEs (\emph{Ordinary Differential Equations}), which is possible in KeymaeraX~\cite{FultonMQVP15,Platzer18} (cf. Example~\ref{ex:cars});

or 

\item proving with probabilistic properties, which is possible in e.g. Rocq~Mathematical Components library~\cite{mathcomp} that  covers substantial fragments of Measure \& Probability Theory~\cite{DBLP:journals/jar/AffeldtC23,AffeldtS24,affeldt2024yet,mca}.
\end{itemize}

\subsection{Challenge 2. Formal Proof Certificates} If we are to use neural network solvers as part of our compilers, as \textbf{Vehicle} or \textbf{CAISAR}~\cite{caisar} do, we need to be sure of their correctness.
Neural network solvers are variants of established proof search methods, such as SMT-solving with bound tightening (Marabou~\cite{katz2019marabou,Wu24}), interval bound propagation ($\alpha\beta$-CROWN~\cite{wang2021beta}), or abstract interpretation (AI2~\cite{GeMiDrTsCgVe18}). 
Unfortunately, even complete neural network solvers are prone to implementation bugs
and numerical imprecision that, in turn, might compromise their soundness and can be maliciously exploited, as was shown in e.g.~\cite{PLNNV25,JiRi21,Zombori2021FoolingAC}. 
 
One could consider verifying neural network solvers directly. However, as Wu et al.~exemplify, a mature neural network verifier is a complex multi-platform library; its direct verification would be close to infeasible~\cite{WuIsZeTaDaKoReAmJuBaHuLaWuZhKoKaBa24}. 
A similar problem was faced by the SMT solvers~\cite{BaDeFo15,BaReKrLaNiNoOzPrViViZoTiBa22,DeBj11}. A solution was found: to produce (and then check) the proof evidence, instead of checking the verifier in its entirety.
The software that checks the proof evidence is called the \emph{proof checker}.

 This idea only works on two conditions: the proof checker should be significantly simpler than the original verifier and it should yield strong guarantees of code correctness. 
For Marabou, the first half of this research agenda was accomplished by Isac et al.~\cite{IsBaZhKa22} with
proof checking for Marabou being reduced to the application of
the \emph{Farkas lemma}~\cite{Va96}, a well-known solvability theorem
for a finite system of linear equations, along with a tree structure that reflects Marabou's verification procedure.  
As the  proof checker in~\cite{IsBaZhKa22} was implemented in C++,  that code itself raised the question of trust 
(e.g., in the precision of floating point calculations or guarantees for implementation soundness).  It thus had no formal guarantee of correctness. 

Therefore, the next step was to establish the necessary guarantees of the checker's correctness, while adhering to the principles of \emph{proof-carrying code}~\cite{Necula97} and avoiding any potential discrepancy between implementation and formalization; and  
 ensuring that the checker itself did not suffer from floating point imprecision. 
The industrial theorem prover Imandra~\cite{passmoreImandraAutomatedReasoning2020} was chosen to for this purpose:
it supports arbitrary precision real arithmetic and can prove properties of programs written in a subset of OCaml. It strikes a balance between interactive and automated proving: i.e., it features strong automation while admitting user interaction with the prover via tactics. 
In Imandra, Desmartin et al. \cite{Des25} obtained the checker implementation that can be safely executed \emph{and} proven correct in the same language.  The result was  the first \emph{certified} proof checker for neural network verification. 
 
 To develop this initial success further, one needs to incorporate similar implementations into languages such as \textbf{Vehicle}, and also extend the range of certified neural network solvers. The new generation of neural network solvers that produce probabilistic, stochastic and/or dynamic guarantees~\cite{AbateGR24,NGL24,mandal2024formallyverifyingdeepreinforcement,manginas2025scalableapproachprobabilisticneurosymbolic} are about to encounter the very same obstacle of missing proof certification. 
However, this time, the mathematical sophistication of formal proofs involved in turning their output into \emph{proof certificates} will require libraries and languages formalising measure/probability theory 
or proofs involving differential equations.  A good hosting language for formalising probabilistic  proof certificates would be an interactive theorem prover with strong Probability theory libraries, such as the Mathematical Components library~\cite{mathcomp,DBLP:journals/jar/AffeldtC23,AffeldtS24,affeldt2024yet,mca}) in Rocq.

\subsection{Challenge 3. Compilation of Verification Properties to Machine Learning Objective Functions}
No implementation of proof-carrying neuro-symbolic programs could be complete without acknowledging that, at the heart of the matter lies the question whether the object we verify -- the neural network $f$ -- actually adheres to the verification conditions we wish to verify in Eq. (1). In Section~\ref{sec:ML} we outlined that this problem boils down to deploying methods that can optimise $f$ to satisfy the desired properties.
Initially, for the robustness properties such as the one given in Eq. (4), various \emph{adversarial training} methods were suggested~\cite{KM18,FischerBDGZV19,ijcai2022p767}.
Generally, the weakest form of property-based training boils down to translating a specification written in a subset of first-order logic into a \emph{loss function}, that serves directly as an optimisation objective within the implementation of a training algorithm. Thus, the training algorithm optimises the neural network to satisfy the desired property.    This translation method is known under the name of \emph{differentiable logic} (or DL) ~\cite{FischerBDGZV19,SlusarzKDSS23,flinkow2024}. \textbf{Vehicle} implements DL as one of its backends~\cite{DBLP:journals/corr/abs-2401-06379} and serves as a prototype of a compiler for neural network property specification languages. 

The remaining challenge in this domain is to establish a coherent theory of the differentiable logics, seen as a sub-family of quantitative logics.  Quantitative logics, i.e.~logics that have semantics in interval domains instead of the Booleans have been studied for decades, and date back to the ideas of Kleene, G\"{o}del, and Łukasiewicz at the start of the 20th century \cite{prooffuzzy}.
They also have been shown to relate to the family of substructural logics, known as \emph{superintuitionistic} logics~\cite{res-lat}.
Fuzzy logics \cite{prooffuzzy}, and the logics of the Lawvere quantale \cite{bacci,dl2} are important examples of quantitative logics. To illustrate, consider a toy syntax with atomic propositions and conjunction, such as
\begin{equation}
\begin{split}
    \Phi \ni \phi &:= A \,|\, \phi \land \phi
\end{split}
\end{equation}
where $A$ is interpreted in a domain $D \subseteq \Ereal$. $D$ varies among logics and restricts the interpretation of connectives. For example,
G\"{o}del logic has an interpretation on $[0,1]$ with conjunctions interpreted as $\min$, and the Lawvere quantale logics would interpret this syntax in the interval $[0, \infty]$ and take multiplication to model conjunction.

Recently, there was a surge of interest in quantitative logics, stimulated by the growing interest in \emph{safer machine learning} \cite{davidad24}. 
Nevertheless, there is one fundamental problem that quantitative logics face in this domain. Many specifications of interest for machine learning are first-order, yet the majority of quantitative logic results concern propositional syntax \cite{ldl,prooffuzzy,bacci}. For example, language in Definition (5) is propositional, while property in Eq. (4) is first-order. Generalizing some sound and complete propositional quantitative logics to first-order logic often comes at a cost of either completeness or function continuity.  
For example, among several known fuzzy logics \cite{ldl,prooffuzzy}, the only first-order extension that is sound and complete involves the Gödel logic, that interprets conjunction as $\min$, disjunction as a $\max$, and universal and existential quantifiers via the infimum and supremum \cite{firstgodel}. However, connectives of this logic are not continuous and therefore not suitable for gradient-descent algorithms.

Recently, a promising solution was proposed by Capucci~\cite{capucci}, interpreting quantifiers as \textit{$p$-means}, a generalization of $p$-norms over a probability space \cite{lpspaces}. This new semantics gives hope that the open problem of finding a suitable approach to quantification in quantitative logics will find its resolution, and  we can soon find a logic that is sound and complete relative to this new quantitative semantics.

Finally, in the context of using quantitative logics for compilation of specifications of neuro-symbolic programs into the machine learning backends, there comes a question of proving their soundness in a theorem prover, in order to obtain a form of a certified machine learning compiler in the style of~\cite{10.1145/3607844,leroyFormalVerificationRealistic2009}.   It has been shown that verifying soundness  of quantitative logics (as well as their so-called  ``geometric properties" assumed by machine learning algorithms: continuity, shadow-lifting, etc.) results in laborious proofs that are error prone \cite{casadio,ldl,taming}. To overcome these challenges, a rigorous formalization of propositional semantics for quantitative (differentiable) logics has been proposed in~\cite{taming}. Extending this formalisation to first-order logics is a non-trivial challenge that is yet to be overcome~\cite{grant,Jairo25}. Moreover, the new semantics proposed by Capucci presents a particular challenge for formal verification, since, unlike the previous formalizations of quantitative logics \cite{taming}, it now also involves results from real analysis and probability. Most notably, it involves formalisation of measure spaces, probability spaces, and Lebesgue integrals, as well as the use of results such as Jensen's  and Hölder's inequalities \cite{inequalities} that require extensions to the existining Mathematical Components Libraries~\cite{mathcomp}. 


\section{Conclusions}
In this paper, we argue for introducing the term \emph{proof-carrying neuro-symbolic code} to refer to implementation of neuro-symbolic programs with strong safety guarantees expressed in the code and maintained by a provably sound compiler. Using three prominent examples of: 
\begin{itemize}
\item a multi-backend neuro-symbolic compiler; 
\item neural network proof certificate checking; and 
\item  formalisation of properties of symbolic loss functions,
\end{itemize}
we illustrated how these novel methods fit with the existing programming language traditions, and how they depart from them or re-interpret them. 

This overview is based on many discussions with my co-authors and colleagues, whose papers were used and cited throughout this paper in order to illustrate more technical points. The interested reader is invited to follow up on these references to learn more. 

\section{Acknowledgments}
The author acknowledges the partial support of the EPSRC grant AISEC: AI Secure and Explainable by Construction (EP/T026960/1),
and support of ARIA: Mathematics for Safe AI grant.  She thanks Matthew Daggitt for providing a nice notation for expressing the essence of neuro-symbolic code with proofs in Section~\ref{sec:nesy}.

%
%
%
\bibliographystyle{splncs04}
\bibliography{bibliography,references}
\end{document}